\documentclass[10pt,journal,compsoc]{IEEEtran}

\usepackage{subcaption}
\newcommand{\etal}{\textit{et al. }}
\usepackage{balance}
\usepackage{breqn}
\usepackage{array}
\usepackage{colortbl,booktabs}
\usepackage{subcaption}
\usepackage{titlesec}
\usepackage{algorithmic}
\usepackage{algorithm}
\usepackage{dblfloatfix}
\usepackage{mathtools,amssymb}
\usepackage{tabularx}
\usepackage{pgfplots} 
\usetikzlibrary{patterns}
\usepackage{graphicx}
\usepackage{multirow}

\ifCLASSOPTIONcompsoc
  \usepackage[nocompress]{cite}
\else
  \usepackage{cite}
\fi

\ifCLASSINFOpdf
\else
\fi

\usepackage{amsfonts}
\usepackage{amssymb}
\usepackage{amsmath}
\usepackage[T1]{fontenc}
\usepackage{mathtools}
\usepackage{breqn}

\def\be{ \begin{equation} }
\def\ee{ \end{equation} }
\def\bea{ \begin{eqnarray} }
\def\eea{ \end{eqnarray} }

\begin{document}

\title{Poisoning Attacks and Defenses in Federated Learning: A Survey}

\author{Subhash Sagar, Chang-Tsun Li, Seng W. Loke, and Jinho Choi
\IEEEcompsocitemizethanks{\IEEEcompsocthanksitem All the authors are with the School of Information Technology, Deakin University, Geelong,
Australia\protect\\
E-mail: \{subhash.sagar,changtsun.li,seng.loke,jinho.choi\}@deakin.edu.au}}

\markboth{IEEE Security and Privacy,~Vol.~, No.~, Dec~2022}%
{Shell \MakeLowercase{\textit{et al.}}: Bare Advanced Demo of IEEEtran.cls for IEEE Computer Society Journals}

\IEEEtitleabstractindextext{%
\begin{abstract}

Federated learning (FL) enables the training of models among distributed clients without compromising the privacy of training datasets, while the invisibility of clients’ datasets and the training process poses a variety of security threats. This survey provides the taxonomy of poisoning attacks and experimental evaluation to discuss the need for robust FL.


\end{abstract}

}

\maketitle

\IEEEdisplaynontitleabstractindextext

\IEEEpeerreviewmaketitle

\section{Introduction}
\label{sec:introduction}

\IEEEPARstart{R}{ecent} advances in distributed machine learning, in particular, federated learning (FL) \cite{Kone15} have paved the way for the next generation of data-driven intelligent applications. FL has emerged as a promising solution to a number of applications to solve data silos while protecting the privacy of data. Since its emergence, FL has been employed in a variety of applications including but not limited to healthcare, crowdsourcing systems, natural language processing (NLP), and the Internet of Things (IoT). 
The primary notion of FL is to generate a collaborative global training model without sharing the actual data distributed over several participating clients. In FL, the participating clients first train local models as well as the shared global model with local data and then send the local updates to the central server in order to update the global model. In this way, the privacy of the client's data is protected from unauthorized access. 

Similar to machine learning, FL is susceptible to several adversarial attacks since both the data and the local training process are controlled by clients.
A few common attacks are inference attacks (targeting the privacy of clients) and poisoning attacks (targeting the training data or model) \cite{9308910}. In poisoning attacks, attackers aim to poison the local training data by injecting the poisoned instances in the training data (i.e., \emph{data poisoning attacks (DPA)}), or directly manipulating the weights of the model updates (i.e., \emph{model poisoning attacks (MPA)}). In essence, the main objective of poisoning attacks is to instigate the global model to predict the attacker-specific outputs on the poisoned inputs, and it is presumed that the main factor of poisoning attacks is the count of malicious clients and the poisoned data. Furthermore, Steinhardt \etal \cite{10.5555/3294996.3295110} suggested that a $3\%$ ratio of malicious clients can reduce the accuracy of the main task by $11\%$. Therefore, it is imperative to study the current state-of-the-art poisoning attacks and defense strategies to mitigate these attacks. 

As of late, a plethora of research has been published in order to comprehensively discuss and understand the notion of FL and its use cases in the real world. 
Most recently, researchers have been exploring the security and privacy threats that limit the exploitation of FL \cite{9308910}
\cite{RODRIGUEZBARROSO2023148}.  
A comprehensive discussion of FL threats is presented in \cite{RODRIGUEZBARROSO2023148} in terms of a taxonomy of attacks and defense methods. The authors also conducted an experimental evaluation in order to draw a conclusion on how to select the suitable method in each category of adversarial attacks. Furthermore, a brief overview of threats to FL is discussed in \cite{9308910} 
and focuses on poisoning and inference attacks in order to comprehend the model behavior in presence of these attacks. 

The generalization of state-of-the-art into various categories is the core of the surveys on security and privacy in FL that have emerged in recent years \cite{9308910}
\cite{RODRIGUEZBARROSO2023148}. The discussed studies' pros and cons with regard to evaluation criteria, including but not limited to the adaptability of attacks these studies can handle, their effectiveness in the presence of backdoor attacks, their use in real-world applications, and their impact on benign clients, are not taken into account by the current literature. In order to get over the aforementioned limitations, this survey presents the most recent state-of-the-art attacks and defense strategies suggested for FL. Additionally, this study lists the advantages and disadvantages of the attacks and defense strategies in relation to a variety of assessment criteria as well as via experimental evaluation. Finally, we have concluded the study by highlighting potential future research directions. 


\section{Preliminaries}

\subsection{Federated Learning}

The notion of FL was first introduced by Google 
and is characterized as a distributed machine learning (ML) paradigm to collaboratively train a global ML model on datasets that are distributed across a number of clients (e.g., mobile devices) while protecting the privacy of clients \cite{Kone15}. 
In essence, there are two different types of parties in FL, i.e., a number of clients and a central server/cloud, wherein each client maintains a local model trained on the local dataset. In contrast, the central server maintains a global model, which is the aggregation of locally trained models. 

In general, FL employs the stochastic gradient descent (SGD) 
to minimize a loss function and the models are updated iteratively (due to the use of SGD). 
In each iteration, there are three stages as follows: (1) selection of clients; (2) local model updates;  and (3) aggregation by the central server. In particular, firstly, the central server selects a number of clients and shares the current global model. Secondly, each client retrains the current global model on the local dataset and sends the latest local model back to the central server.
Finally, the central server aggregates all the local models in order to obtain the latest global model and such aggregation schemes include mean aggregation, byzantine-robust aggregation, etc \cite{RODRIGUEZBARROSO2023148}.

\subsection{Poisoning Attacks}

The term `poisoning attack' in FL refers to an attack that involves attackers intentionally tampering with the training data or model parameters to manipulate the model aggregation in order to disrupt the integrity and availability of the global learning model. 
In FL, because the participants' data and the training processes are invisible to the central server, poisoning attacks might be easy to carry out by some clients.

\begin{figure*}
     \centering
     \begin{subfigure}[b]{0.45\textwidth}
        \centering
        \includegraphics[width=0.85\textwidth]{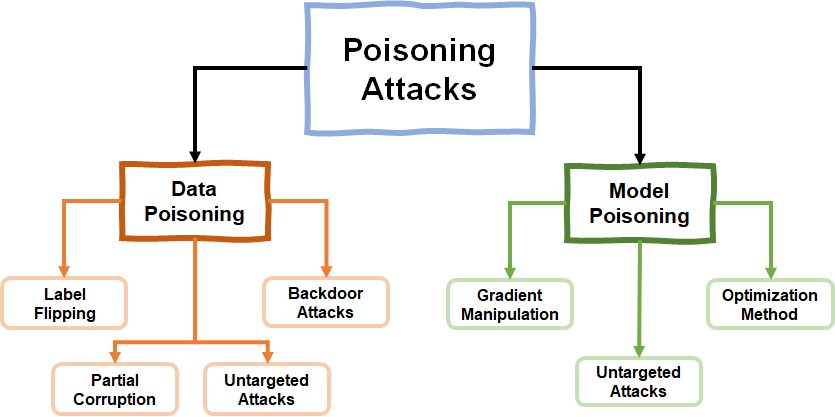}
        \caption{Taxonomy of Poisoning Attacks in Federated Learning}
        \label{fig:fl_pa_tax}
     \end{subfigure}
     \hfill
     \begin{subfigure}[b]{0.45\textwidth}
        \centering
        \includegraphics[width=0.75\textwidth]{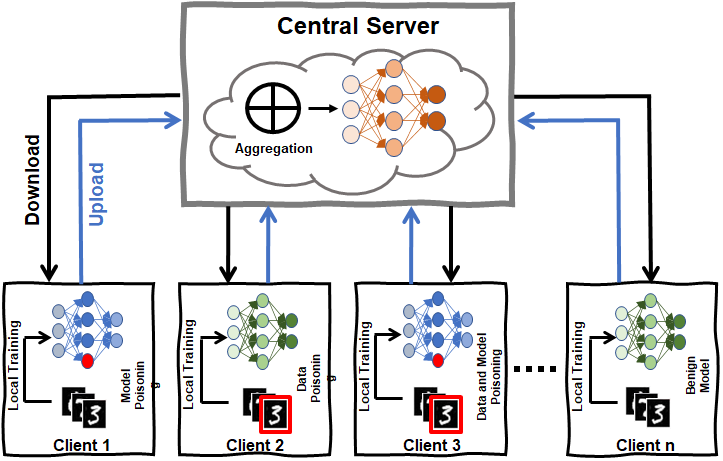}
        \caption{Illustration of Poisoning Attacks in Federated Learning}
        \label{fig:tax_pa_fl}
     \end{subfigure}
        \caption{Taxonomy and illustration of poisoning attacks}
        \label{fig:three graphs}
\end{figure*}

Poisoning attacks can be divided into two types, i.e., DPA and MPA, or can be divided into targeted and untargeted attacks \cite{RODRIGUEZBARROSO2023148}. We have considered the division in terms of data and poisoning attacks and a taxonomy of these attacks is illustrated in Figure~\ref{fig:fl_pa_tax}. The location of poisoning attacks may vary and it can be on the client's side, communication channel, or sometimes on the central server. Nevertheless, the common poisoning attacks usually take place on the client side (i.e., modifying the data, models, or both) as illustrated in Figure~\ref{fig:tax_pa_fl}.

\subsubsection{Data Poisoning Attacks} 

In DPA, it is presumed that the attackers have access to the training data of at least one client and are able to alter it. DPA can be further classified in terms of the characteristic of the poisoning, and we have categorized them into the following types of attacks. 

\noindent \emph{Label-Flipping Attacks:} In this type of attack, adversaries with access to the training data alter the labels of a portion of the data (e.g., flipping label $4$ to $1$, or vice versa) while preserving the remaining content in order to manipulate the FL models. Label flipping can be either targeted (i.e., flipping the targeted label) or untargeted (i.e., random flipping) \cite{10.1007/978-3-030-58951-6_24}. 

\noindent \emph{Poisoning Sample Attacks:} In this type of DPA, the attackers modify a portion of training data, i.e., by inserting modified patterns in data samples or adding unification noise. Recent studies suggest the use of generative adversarial nets to generate positioned patterns in order to maximize targeted attacks and evade defense approaches \cite{10.1007/978-3-030-58951-6_24}. 

\noindent \emph{Backdoor Attacks:} The idea of backdoor attacks is to degrade the performance of a subtask while preserving the performance of the main task. These attacks inject triggers into the training data of one or more clients' to poison the global model. Since these attacks do not affect the overall performance, it is difficult to mitigate. In essence, backdoor attacks do not pose threats to the FL's main task. However, the integrity and the infrastructure of FL are vulnerable and can be considered a security threat as the prediction of test samples can be controlled by attackers \cite{Xie2020DBA}\cite{NEURIPS2020_b8ffa41d}. 

\noindent \emph{Untargeted Attacks:} The untargeted attacks aim at disrupting the performance of the FL model. One of the well-known scenarios is the \emph{byzantine attack}, wherein an adversary shares a randomly generated model trained over modified training data. The untargeted attacks can be further divided into the following two types, namely \emph{disruption untargeted} and \emph{exploitation untargeted} \cite{RODRIGUEZBARROSO2023148}. With disruption untargeted attacks, the attackers corrupt the FL model in order to disrupt the convergence of the training process. With exploitation untargeted attacks, the attackers utilize the poisoning attacks in order to exploit the FL framework for malicious purposes by mimicking as a benign client .

\subsubsection{Model Poisoning Attacks} 

MPA aims to manipulate the training processes of the local FL models by directly poisoning the model updates sent by the clients to the central server. Furthermore,  DPA can be considered to be a subset of MPA as they eventually lead to MPA, and in general, MPA attempts to directly modify  local model weights. A number of variants fall under the category of model poisoning and are discussed as follows. 

\noindent \emph{Gradient Manipulation:} These types of attacks \cite{pmlr-v108-bagdasaryan20a} are based on generating random weights having a similar dimension as that of the original model weights. These random weights are then employed to manipulate the local model gradient and compromise the global model. 

\noindent \emph{Optimization Methods:} Optimization methods \cite{10.5555/3489212.3489304} are utilized to maximize the performance of poisoning attacks, specifically, backdoor attacks, wherein the optimization helps in minimizing the difference between the original and the poisoned model from the last round in order to make it difficult for the attacks to be mitigated. Hence, the attackers force the main model to output the attacker-specified labels.  

\noindent \emph{Untargeted Attacks:} The untargeted attacks are to lower the main task accuracy of the global model \cite{10.1145/3534678.3539231}. Moreover, the attackers in these types of attacks quantify the original local model updates, and then alter their genuine local model updates so that the tainted global model updates significantly diverge from the original ones. In essence, these attacks aim at limiting the availability of the global model.  


\section{Poisoning Attacks and Defense Strategies}

This section discusses existing state-of-the-art for poisoning attacks and defense strategies in FL.  

\subsection{Attacks Strategies}

In FL, poisoning attacks work by corrupting the data of one or more clients or the locally trained models in an effort to interfere with the global model. For DPA, it is assumed that the attacker has the access to the data of at least one client and the capacity to alter it, wherein the poisoned data is an amalgam of cleaned (original labels) and modified data. Furthermore, the MPA directly poison the local model updates sent by the participating clients to the global server by modifying the weights of the local models. 

With the assumption that the FL participants may change the raw data on their devices, the authors of \cite{10.1007/978-3-030-58951-6_24} proposed a DPA model in terms of label flipping attacks. Under this assumption, a complex neural network model is employed to investigate the FL application in the presence of label-flipping attacks. Furthermore, the efficacy of the attacks is shown with respect to varying percentages of malevolent participants using two well-known image datasets, \emph{CIFAR-10} and \emph{Fashion-MNIST}. It can be observed that the attack is effective even with a small ratio of malicious participants and can also be targeted, i.e., having more impact on a subset of targeted data labels.

Xie et al. suggested a distributed backdoor attack, called DBA \cite{Xie2020DBA}, which employed a composite global trigger to perform the attacks as opposed to the straightforward label-flipping attack \cite{10.1007/978-3-030-58951-6_24} carried out by several attackers. The local models are trained on a poisoned dataset after each adversary picks a unique local trigger to poison the training dataset. The server then receives the updates from the attackers for model aggregation. The attackers employ the local triggers to create a global trigger at the inference step rather than using them directly. It is demonstrated that the global trigger has the highest attack success rate of any local trigger, even if it never appears in the training phase.

Furthermore, \cite{NEURIPS2020_b8ffa41d} theoretically demonstrated that backdoor attacks are inevitable in the presence of adversarial samples in FL and intuitively proposed a new family of backdoor attacks, known as edge-case backdoor to manipulate the model on the input resides on the tail of the distribution. Therefore, this makes it difficult for the defense strategies to detect such attacks as the attacker’s access is restricted to developing an untrained feature map instead of a fully trained model.

In \cite{pmlr-v108-bagdasaryan20a}, the vulnerability of FL towards a MPA is discussed. The MPA can be performed by using model replacement, wherein one or more malicious clients attempt to replace a benign model with a malicious one, leading the model to misclassify future inputs. Moreover, the study discusses the countermeasures and concludes that the attackers can easily bypass such measures. The main idea of FL is to take advantage of the diversity of the clients in terms of non-iid training data, including uncommon or low-quality data. However, using secure aggregation by the central server to filter out anomalous contributions runs counter to this idea.

In contrast to \cite{pmlr-v108-bagdasaryan20a}, Bhagoji \etal proposed \cite{bhagoji2019analyzing} to carry out the attack even when the global model is not yet converged. They specifically nurture the malicious updates for $lambda$ times to counteract the benefits of benign clients. Two stealth metrics that the server might examine were suggested. The first is that the server may determine whether the attacker's update aids in model training, i.e., enhances the performance of the overall model. The server may also examine if the submitted update differs from earlier updates, which is the second possibility. They then improved the loss function based on those two covert metrics to prevent anomaly identification in order to increase the attack's robustness. Finally, the effectiveness of the model is evaluated with a number of assumptions (single-shot, repeated, and pixel-pattern attack). It was observed that the model can achieve $100\%$ accuracy on an attacker's task in just a single training round.   


\renewcommand{\arraystretch}{1.5}
\begin{table*}[h]
  \begin{center}
    \caption{Comparison of Attacks Strategies}
    \label{tab:FL_attacks}
    \begin{tabular}{|>{\centering\arraybackslash}m{1.5cm}| >{\centering\arraybackslash}m{2.5cm}| >{\centering\arraybackslash}m{2cm}| >{\centering\arraybackslash}m{2cm}| >{\centering\arraybackslash}m{2cm}| >{\centering\arraybackslash}m{3.5cm}|}
     \hline
      \textbf{Literature} & \textbf{Attacks Type} & \textbf{Training Set Type} & \textbf{Adaptive Attacks} & \textbf{Backdoor Attacks} &  \textbf{Application} \\ 
      \hline 
      
      \cite{10.1007/978-3-030-58951-6_24} & Data Poisoning & iid & No & No & Image Classification \\
      \hline
           
      
      \cite{Xie2020DBA} & Data Poisoning & Non-iid & No & Yes & Image Classification\\ \hline
      
      \cite{NEURIPS2020_b8ffa41d} & Data Poisoning & iid & Yes & Yes & Image Classification, Text Prediction and Sentiment Analysis\\ \hline
      
      \cite{pmlr-v108-bagdasaryan20a} & Model Poisoning & Non-iid & Yes & Yes & Image Classification and Word Prediction \\ \hline
      
      \cite{bhagoji2019analyzing} & Model Poisoning & iid & Yes & Yes & Image Classification and Census Income Prediction\\ \hline
    
      \cite{10.5555/3489212.3489304} & Model Poisoning & Non-iid & Yes & Yes & Image Classification and Breast Cancer Detection\\ 
      \hline

      \multicolumn{2}{|c|}{\textbf{Attack Types}} &
      \multicolumn{2}{|c|}{\textbf{Training Set Types}} & \multicolumn{2}{|c|}{\textbf{Adaptive Attacks}} \\ \hline
      \multicolumn{2}{|c|}{Data Poisoning} &
      \multicolumn{2}{|c|}{iid} &
      \multicolumn{2}{|c|}{Yes} \\  
      \multicolumn{2}{|c|}{Model Poisoning} &
      \multicolumn{2}{|c|}{Non-iid} &
      \multicolumn{2}{|c|}{No} \\  
      \hline
      \multicolumn{3}{|c|}{\textbf{Backdoor Attacks}} &
      \multicolumn{3}{|c|}{\textbf{Target Application}}   \\ \hline
      \multicolumn{3}{|c|}{Yes} &
      \multicolumn{3}{|c|} {Yes}  \\
      \multicolumn{3}{|c|}{No} &
      \multicolumn{3}{|c|} {No Specific Application}  \\
      \hline

    \end{tabular}
  \end{center}
\end{table*}

\subsection{Defense Approaches}
A number of defense strategies are present in the literature, this study divides the state-of-the-art into three defense types, namely \emph{anomaly detection}, \emph{robust aggregation}, and \emph{perturbation mechanism}. 

\noindent \textbf{Anomaly Detection:} In general, anomaly detection is understood as the identification of rare events or observations significantly different from the notion of well-defined data or activities. A defense algorithm to mitigate the label-flipping attacks is proposed in \cite{10.1007/978-3-030-58951-6_24}, wherein the aggregator employs an array to store the subset of parameters from the final layers of the deep neural network (DNN) architecture in each round. The list of stored parameters is then fed into principle component analysis to cluster into honest and malicious participants and accordingly limit the participation of malicious clients in model training. Similarly, 
 A two-phase defense mechanism, known as LoMar is proposed in \cite{9650669}, wherein the kernel density approximation is employed in order to score the local model update over the neighbouring update. The measured score is utilized to cluster participants with similar characteristics and then employ an outlier detection technique to mitigate the poisoning attack. 
 

\noindent \textbf{Robust Aggregation:} Aggregation mechanisms are utilized to mitigate poisoning attacks during the training phase in contrast to anomaly detection, and a few robust aggregation mechanisms are Krum, mean, trimmed mean, etc. Furthermore, most of these methodologies assume the iid data, therefore, fail to mitigate against current poisoning attacks for non-iid scenarios. 

A feedback-based defense strategy, known as \emph{BaFFle}, is put forward in \cite{9546463} to eliminate the backdoored model updates. \emph{BaFFle} employs each participant to validate the global model for each round of FL by quantifying the validation function with their own private data to report whether the model is backdoored or not to the central server. The validation function compares the misclassification rate of a distinct class with the previous global model. Then, the central server utilizes the feedback via the validation function to decide whether to accept or reject the global model update based on the misclassification rate. 

The discussed defense strategies work fine in the presence of a small number of malicious participants. The performance of these strategies degrades with an increase in malicious participants. Zhang \etal \cite{10.1145/3534678.3539231} proposed the malicious client detection algorithm, named \emph{FLDetector}, wherein the idea is to detect the malicious participants and eliminate them from the global model updates. In \emph{FLDetector}, the central server predicts the participants' model updates in each iteration based on historical updates and flags the participant as malicious if there is an inconsistency between the current update and the predicted one in multiple iterations. The performance evaluation illustrates promising results when compared with the current state-of-the-art. Nevertheless, the impact of eliminating the malicious participants on the main task accuracy is not discussed. 

\noindent \textbf{Perturbation Mechanism:} The perturbation mechanism in mathematics represents a method of solving a problem by comparing it with the known solution. One of the well-known perturbation mechanisms is differential privacy, 
which has been extensively employed in FL 
to improve the performance of outlier detection. A low-complexity defense approach is proposed in \cite{sun2019can} in order to mitigate the backdoored model updates via weight clipping and noise injection. However, this approach fails to mitigate the untargeted backdoored attacks that do not rely on model weight modification. In essence, the weight clipping method may negatively influence the change in the weights of benign participants' model updates. Most recently, to overcome the limitation of current differential privacy-based strategies that deteriorate the performance of benign participants, a novel technique, called (\emph{FLAME}) \cite{280048}, is proposed. \emph{FLAME} estimates the amount of noise to be injected to mitigate the targeted poisoning attacks with minimal impact on benign participants. This technique employs the clustering and weight-clipping methods to eliminate outliers in the participants' updates and then the estimated noise is added to obtained parameters to mitigate possible attacks. The experimental evaluation demonstrates that \emph{FLAME} is effective in defending against poisoning attacks when compared with the state-of-the-art. However, it requires extensive modification in the current FL framework. 

\begin{figure*}[t]
     \centering
     \begin{subfigure}[b]{0.235\textwidth}
         \centering
         \includegraphics[width=\textwidth]{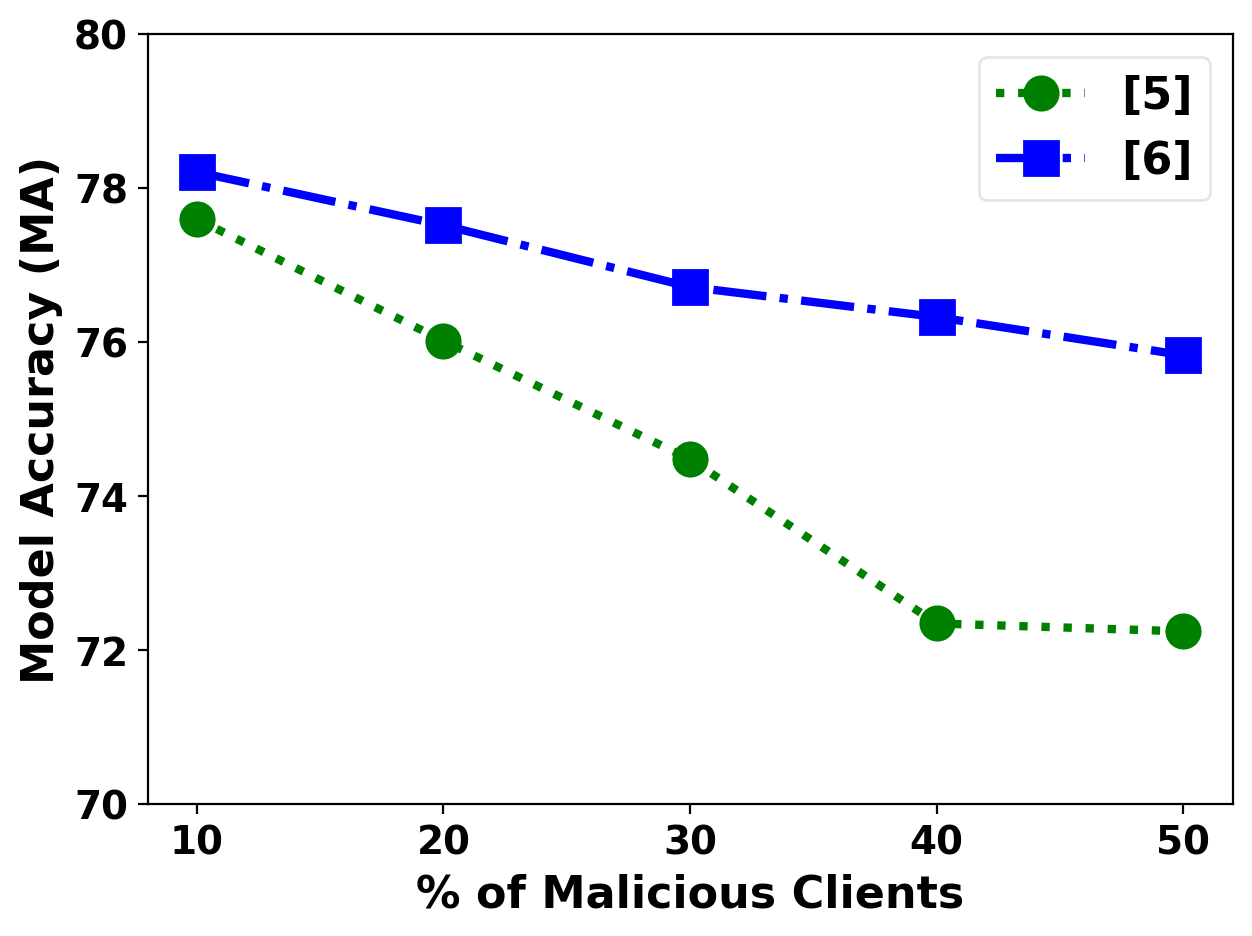}
         \caption{Model Accuracy - DPA}
         \label{fig:acc_model_dpa}
     \end{subfigure}
     \hfill
     \begin{subfigure}[b]{0.235\textwidth}
         \centering
         \includegraphics[width=\textwidth]{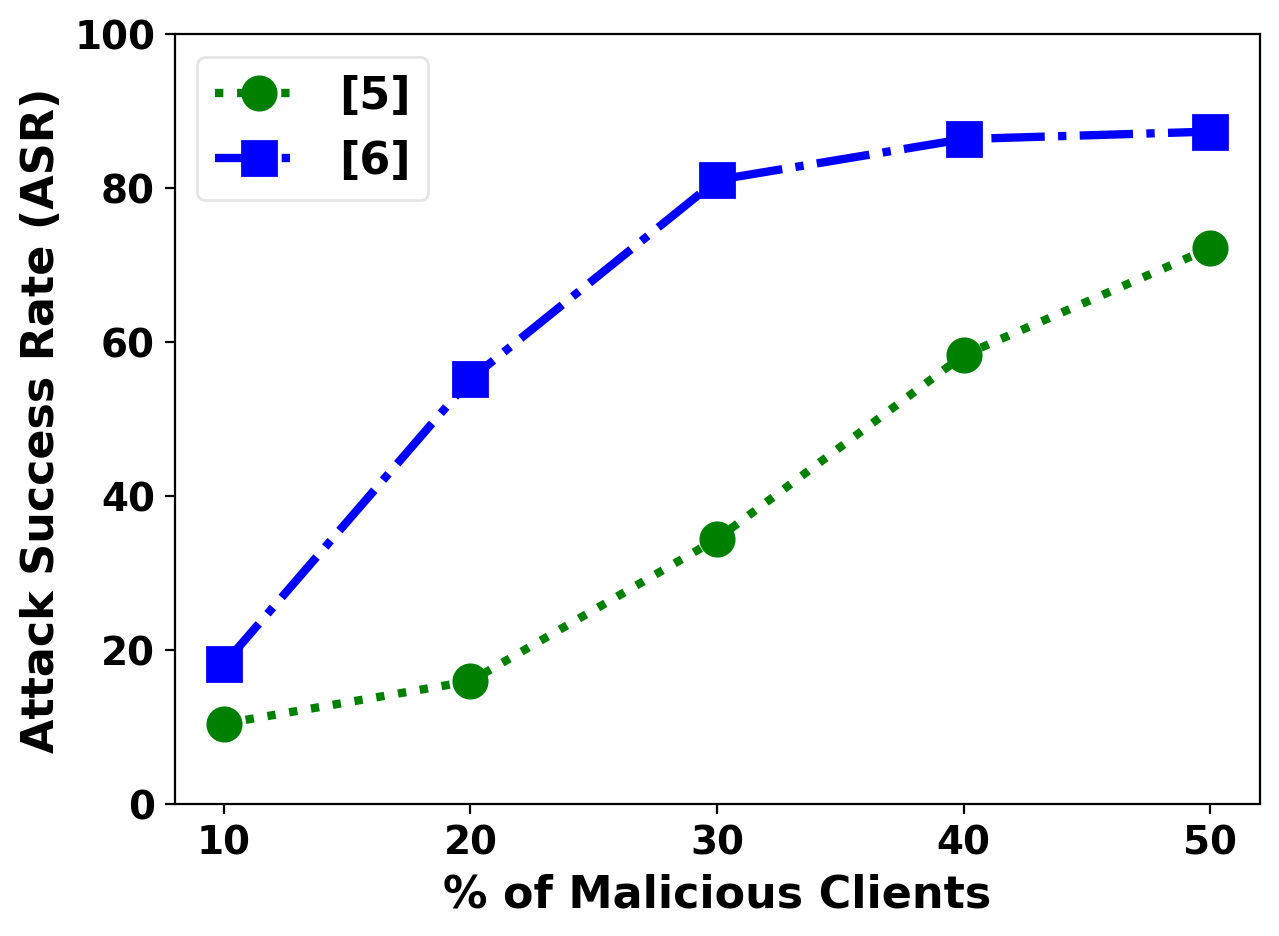}
         \caption{Attack Success Rate - DPA}
         \label{fig:succ_attack_dpa}
     \end{subfigure}
     \hfill
     \begin{subfigure}[b]{0.235\textwidth}
         \centering
         \includegraphics[width=\textwidth]{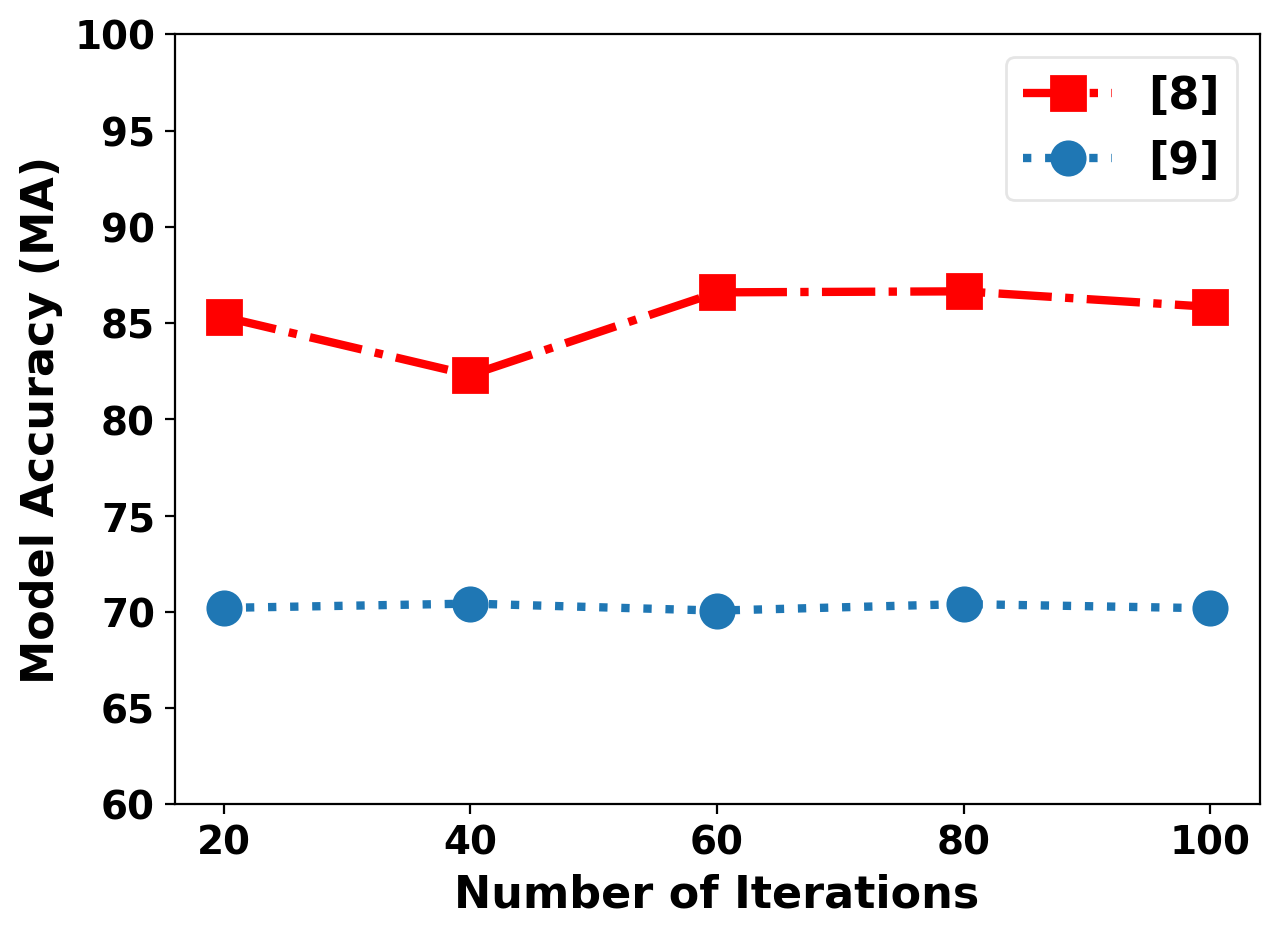}
         \caption{Model Accuracy - MPA}
         \label{fig:acc_model_mpa}
     \end{subfigure}
     \hfill
     \begin{subfigure}[b]{0.235\textwidth}
         \centering
         \includegraphics[width=\textwidth]{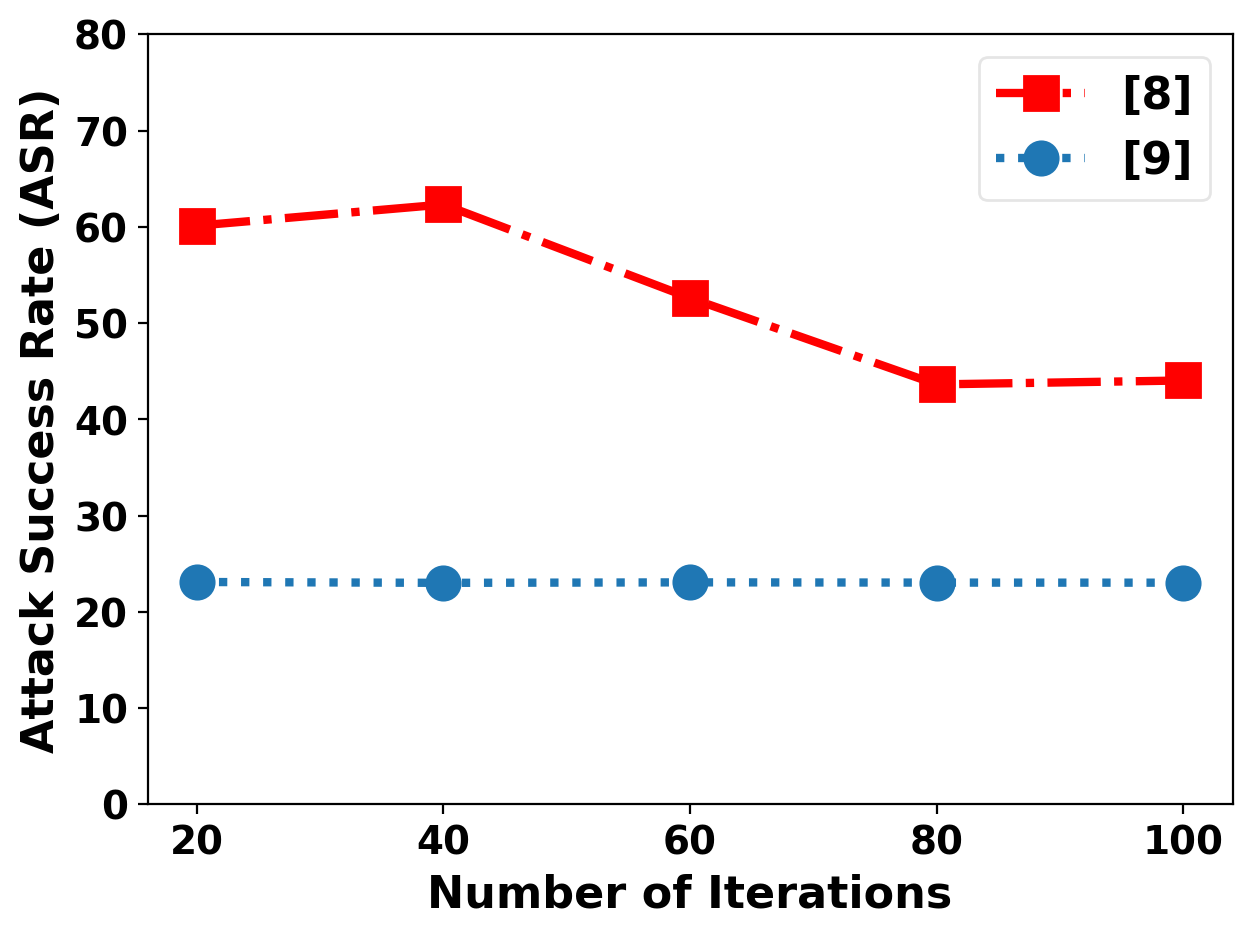}
         \caption{Attack Success Rate - MPA}
         \label{fig:succ_attack_mpa}
     \end{subfigure}     
        \caption{Poisoning Attacks - Model Accuracy and Attack Success Rate on CIFAR-10 w.r.t. varying malicious clients for DPA and varying iterations for MPA}
        \label{fig:poisoning_attacks}
\end{figure*}

\section{Analysis and Evaluation}

This section presents the analysis and performance evaluation of poisoning attacks and the defense strategies. 

\subsection{Comparison and Evaluation of Attack Strategies}

Firstly, we compare attack strategies with the following fours characteristics (Table \ref{tab:FL_attacks})

\noindent \emph{Training Set Types:} FL is devised to employ clients’ data while protecting the privacy, and there are two types of data in general, i.e., independent and identically distributed (iid) and non-independent and identically distributed (non-iid). Here, iid data samples are statistically identical and drawn from the same probability distribution, wherein each data sample is seen as an independent event, while non-iid data samples are not statistically identical and include complexities beyond iid, including but not limited to relationships, heterogeneity, dynamic data over time, sampling, etc. 
In general, most of the real-life datasets are non-iid. However, the current notion of analytics and machine learning methods often are based on the assumption that datasets are iid. Therefore, it is imperative to assess the proposed solutions on training set type in terms of their suitability for real-world applications.

As can be seen in Table \ref{tab:FL_attacks}, the majority of the attack strategies utilize iid datasets for DPA (except \cite{Xie2020DBA}) in order to distinguish between the benign and malicious updates \cite{10.1007/978-3-030-58951-6_24}
\cite{NEURIPS2020_b8ffa41d}. In contrast, the majority of MPA use datasets that are non-iid, making them beneficial in real-life FL environments. 

\noindent \emph{Adaptive Attacks:} An adaptive attack represents the variation in the attack scenario of the attacker, e.g., each round, an attacker modifies its strategies to attack the data or model by learning the defense strategies of the FL system. A number of state-of-the-art attack approaches consider adaptive attacks 
\cite{NEURIPS2020_b8ffa41d}\cite{pmlr-v108-bagdasaryan20a}\cite{10.5555/3489212.3489304} to disrupt the functionality of federated learning systems via poisoning attacks (i.e., data and model) from the intelligent malicious clients.

\noindent \emph{Backdoor Attacks:} A backdoor attack aims to mislead the model to perform abnormally on data samples stamped with a backdoor trigger to misclassify specific target labels in terms of targeted backdoor but behave normally on all other samples. Moreover, the untargeted backdoor attacks aim to degrade the overall accuracy of the global model \cite{RODRIGUEZBARROSO2023148}. The majority of the studies discussed in this research consider the backdoor attacks to mislead the backdoor model by injecting local triggers to poison the model (except \cite{10.1007/978-3-030-58951-6_24})

\noindent\emph{Application:} It is also imperative to identify the practicality of the proposed attacks model in terms of a specific application in order to devise the conclusion on the suitability of the current state-of-the-art. As can be seen from the table that most of the proposed attack strategies focus on poisoning attacks on image classification. 

 The experimental evaluation in terms of overall model accuracy (MA) and attack success rate (ASR) is carried out for a few selected attacks models \cite{10.1007/978-3-030-58951-6_24}\cite{Xie2020DBA}\cite{pmlr-v108-bagdasaryan20a}\cite{bhagoji2019analyzing} as depicted in Figure \ref{fig:poisoning_attacks} for both DPA and MPA. MA measures the model accuracy on benign samples, whereas, ASR represents the probability (success rate) of attacks in misclassifying the target labels. 

We have trained the global model on CIFAR-10 for attack scenarios and MNIST for defense strategies with a total of 100 participants and considered the common simulation scenarios among all the compared models. For DPA, we train the global model with a varying number of malicious participants. For the MPA, the global model is trained with a varying number of iterations with a poisoning rate of $10\%$. 

The results for DPA are shown in Figure \ref{fig:poisoning_attacks}a and \ref{fig:poisoning_attacks}b. It can be seen that \cite{Xie2020DBA} can achieve higher ASR (Figure \ref{fig:acc_model_dpa}) than \cite{10.1007/978-3-030-58951-6_24} and also maintains high MA (Figure \ref{fig:succ_attack_dpa}). The reason for lower ASR and lower MA for \cite{10.1007/978-3-030-58951-6_24} is the use of complex CIFAR-10 datasets and the increase in the malicious ratio. Moreover, the results for MPA are depicted in Figure \ref{fig:poisoning_attacks}c and \ref{fig:poisoning_attacks}d. It can be seen that the ASR for \cite{pmlr-v108-bagdasaryan20a} decreases with an increase in the iterations as the benign participants dilute the impact of backdoor. However, the ASR and MA of \cite{bhagoji2019analyzing} remains stable with the iterations. 

In general, the current studies are well-suited to poisoning attack scenarios with the fixed assumption. Nevertheless, the predefined conditions in the state-of-the-art attack strategies may not be suitable for real-world applications with dynamic characteristics. 

\begin{table*}[h]
  \begin{center}
    \caption{Comparison of Defense Strategies}
    \label{tab:FL_defense}
    \begin{tabular}{|>{\centering\arraybackslash}m{1.2cm}| >{\centering\arraybackslash}m{3.3cm}| >{\centering\arraybackslash}m{2.25cm}|
    >{\centering\arraybackslash}m{2.5cm}| >{\centering\arraybackslash}m{1.9cm}|
    >{\centering\arraybackslash}m{1.8cm}|>{\centering\arraybackslash}m{2cm}|}
     \hline
      \textbf{Literature} & \textbf{Defense Type} & \textbf{Attack Type} & \textbf{Defense Target} &\textbf{Training Time Defenses} & \textbf{Secure Aggregation} & \textbf{Effect on Benign Clients}\\ 
      \hline 
      \cite{10.1007/978-3-030-58951-6_24} & Anomaly Detection & Data Poisoning & Label Flipping & Yes & No & No\\ \hline

      \cite{9650669} & Anomaly Detection & Data Poisoning & Label Flipping & Yes & No & No\\ \hline


      \cite{9546463} & Robust Aggregation & Model Poisoning & Backdoor Attacks & No & Yes & No\\ \hline
 
      \cite{10.1145/3534678.3539231} & Robust Aggregation & Model Poisoning & Untargeted Attacks & Yes & Yes & Yes\\ \hline
  
      \cite{sun2019can} & Perturbation Mechanism & Model Poisoning & Backdoor Attacks & Yes & Yes & Yes\\ \hline
      
      \cite{280048} & Perturbation Mechanism & Model Poisoning & Backdoor Attacks & Yes & No & No\\
      \hline   
      
      \multicolumn{3}{|c|}{\textbf{Defense Types}} &
      \multicolumn{2}{|c|}{\textbf{Attack Type}} & \multicolumn{2}{|c|}{\textbf{Defense Target}} \\ \hline
      \multicolumn{3}{|c|}{Anomaly Detection} &
      \multicolumn{2}{|c|}{Data Poisoning} &
      \multicolumn{2}{|c|}{Label Flipping} \\  
      \multicolumn{3}{|c|}{Perturbation Mechanism} &
      \multicolumn{2}{|c|}{Model Poisoning} &
      \multicolumn{2}{|c|}{Backdoor Attacks} \\  
      \multicolumn{3}{|c|}{Robust Aggregation} &
      \multicolumn{2}{|c|} {-} &
      \multicolumn{2}{|c|}{Untargeted Attacks} \\
      \hline
      \multicolumn{3}{|c|}{\textbf{Training Type Defenses}} &
      \multicolumn{2}{|c|}{\textbf{Secure Aggregation}} & 
      \multicolumn{2}{|c|}{\textbf{Influence on Benign Clients}} \\ \hline
      \multicolumn{3}{|c|}{Yes} &
      \multicolumn{2}{|c|} {Yes} & \multicolumn{2}{|c|} {Yes} \\  \multicolumn{3}{|c|} {Yes} &  
      \multicolumn{2}{|c|}{Np} & \multicolumn{2}{|c|} {No}
\\
      \hline
    \end{tabular}
  \end{center}
\end{table*}

\subsection{Comparison and Evaluation of Defense Strategies}

Table \ref{tab:FL_defense} illustrates the comparison of defense strategies by employing a number of aspects discussed below.

\noindent \emph{Defense Types:} A variety of defense strategies are suggested to lessen the effects of poisoning attacks and it is imperative to compare the effectiveness of each of these strategies since they each have merits and drawbacks. The defense strategies described in the literature have generally been grouped into three categories: \emph{anomaly detection}, \emph{perturbation mechanism}, and \emph{robust aggregation}. In essence, the literature uses either the robust aggregation or perturbation method for the MPA \cite{10.1145/3534678.3539231}\cite{280048}\cite{9546463}, but the defensive strategies for DPA include anomaly detection as the defense type to mitigate these attacks \cite{10.1007/978-3-030-58951-6_24}
\cite{9650669}.

\noindent \emph{Defense Target:} This parameter compares the existing literature in terms of types of targeted attacks mitigated by the current state-of-the-art (e.g., label-flipping attacks, backdoor attacks, untargeted attacks, etc). The label-flipping attack is the common DPA mitigated by the present state-of-the-art \cite{10.1007/978-3-030-58951-6_24}\cite{9650669},
and in contrast, backdoor attacks and untargeted attacks are the defense target for the MPA \cite{10.1145/3534678.3539231}\cite{280048}\cite{9546463}. 

\noindent \emph{Training Time Defense:} Most of the proposed solutions are effective in mitigating poisoning attacks during the training time period as the global model does not have access to client's data, therefore, the submitted local model is evaluated in terms of its reliability. However, a few studies consider applying the defense strategies on the model submitted by the clients after training \cite{9546463}. In these studies, the defender prunes the dormant neurons after the training step via the submitted pruning sequence from the client. 

\noindent \emph{Secure Aggregation:} 
In order to prevent the inspection of confidential model updates, a number of FL works suggest utilizing secure multi-party computation (MPC). MPC is a subbranch of cryptography, wherein the goal is to create a method for multiple parties to jointly compute a function over their inputs 
while preserving privacy. Thus, the comparison of current studies in terms of compatibility with secure aggregation is essential. Most of the FL systems employ MPC to secure submitted confidential models from the clients. However, a few defense strategies \cite{10.1007/978-3-030-58951-6_24}\cite{280048}  rely on detecting the local model updates and are not compatible with secure aggregation.

\noindent \emph{Effect on Benign Clients:} It is essential for the defense strategies to not only detect malicious clients but also have minimal effect on benign clients. A number of studies have a negative effect on benign clients. For example, the authors in \cite{10.1145/3534678.3539231} utilize the suspicious score by comparing the predicted and updated model to detect malicious clients. This score can cause ill-effect on benign clients if the score is closer to the boundary of the suspicious score. The performance of the benign clients is impacted by the differential privacy mechanism in \cite{sun2019can} because the clipping factor will alter the benign updates.

\begin{figure}
     \centering
     \begin{subfigure}[b]{0.235\textwidth}
         \centering
         \includegraphics[width=\textwidth]{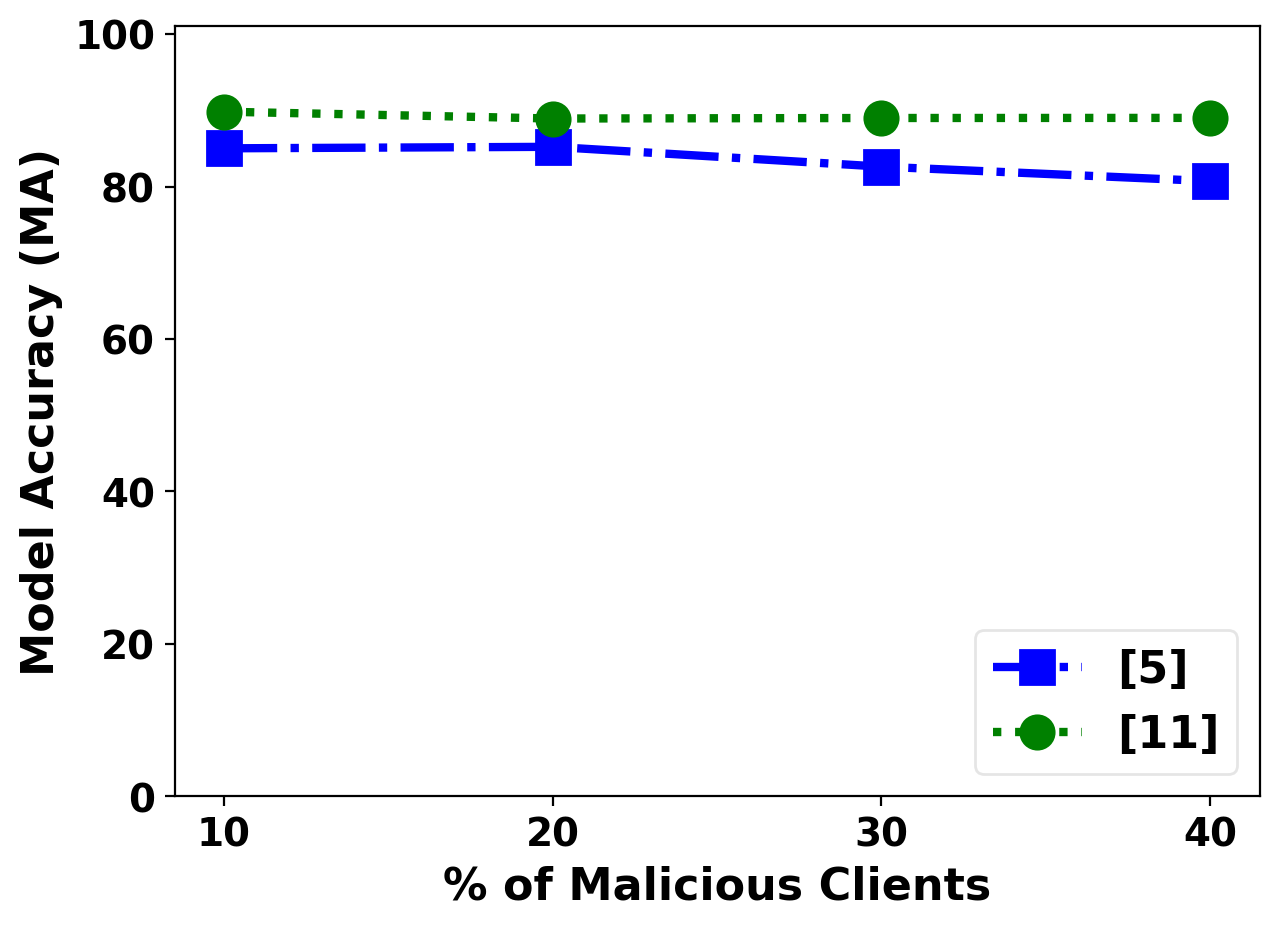}
         \caption{Defense Against DPA}
         \label{fig:acc_def_dpa}
     \end{subfigure}
     \begin{subfigure}[b]{0.235\textwidth}
         \centering
         \includegraphics[width=\textwidth]{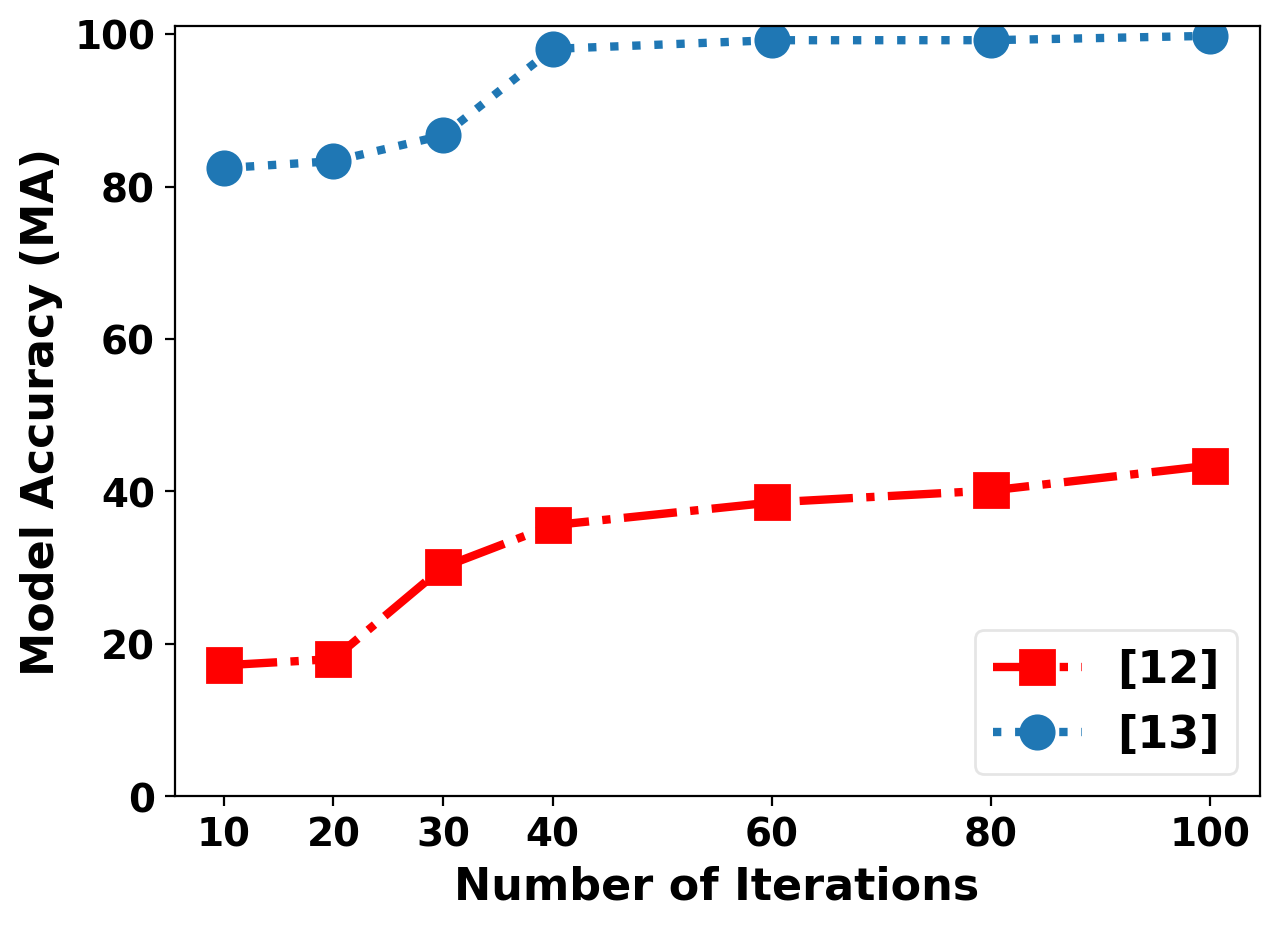}
         \caption{Defense Against MPA}
         \label{fig:acc_def_mpa}
     \end{subfigure}
     \caption{Model Accuracy w.r.t. varying percentage of Malicious Clients and Number of Iterations}
     \label{fig:defense_strat}
\end{figure}

In Figure \ref{fig:defense_strat}, we have compared the four defense strategies studied in \cite{10.1007/978-3-030-58951-6_24}\cite{9650669}\cite{9546463}\cite{10.1145/3534678.3539231} in terms of MA with respect to varying percentages of malicious participants  the number of iterations for DPA and MPA, respectively.  As depicted in Figure \ref{fig:acc_def_dpa}, the MA for \cite{10.1007/978-3-030-58951-6_24} decreases as the number of malicious participants increases, and maintains the accuracy of around $80\%$. Nevertheless, the accuracy of \cite{9650669} remains stable with an average MA of $85\%$. Furthermore, the MA for defense strategies for MPA increases as the number of iterations increases for both \cite{10.1145/3534678.3539231} and \cite{9546463}. Nonetheless, the convergence time in terms of defense mechanism for \cite{10.1145/3534678.3539231} is better than \cite{9546463}.

As a whole, the current research on defense strategies for poisoning attacks is well-defined for the specific scenarios. However, intelligent malicious clients can still learn the behaviour of these defense strategies and poison the FL systems. 

\section{Future Research Directions}

As the notion of the FL paradigm has been widely explored in recent years, adversarial attacks and defenses for FL are also investigated. Nevertheless, there are still numerous research challenges that need the attention of researchers. We have identified a number of such research challenges in this section. 

Firstly, the mathematical formulation for different types of poisoning attacks is not lacking. For example, in data poisoning, the label-flipping attack depends on the number of malicious clients in the data and it is not deterministic (i.e., no mathematical exposition of this phenomenon). Furthermore, most of the attack strategies are inclined towards data with similar feature space, and different samples. However, the behaviour of these attacks will vary with data having different feature space (i.e., vertical FL) as the label information is usually  hidden in this type of FL. In general, proposing the mathematical exposition of attacks and designing attack strategies suitable for vertical FL are potential research directions. 

Secondly, defense strategies proposed in the literature somehow have vulnerabilities that can be transformed into another adversarial attack, hence, the study of these vulnerabilities and the solutions to mitigate such attacks is imperative and challenging. Furthermore, most defense strategies have trade-offs between attack prevention and the overall performance of the original task.
For example, a solution based on differential privacy tends to add noise in order to preserve privacy, which however impairs the performance of the global model. In addition, the client filtering-based defense solutions sometimes lead to filtering out more clients and thus, losing a large portion of data for the aggregation process. In general, it is imperative to consider an optimal defense strategy that can filter out all the attacks while maintaining the original task's performance. The solution must not compromise the privacy of the clients while preventing adversarial attacks. 

Thirdly, the importance of the types of training set cannot be understated since it is difficult to distinguish between malicious and benign clients with various data samples that are important for the learning process in the case of non-iid distribution of data. Employing an anomaly detection algorithm suitable for non-iid distribution or those that do not rely on data distribution 
is the typical solution to these situations. Nevertheless, it is a problem that requires attention because the methods mentioned above still struggle to identify malicious clients in situations with highly skewed data distribution.

Finally, keeping track of adversaries is one of the key challenges since clients in the FL system are free to leave and rejoin the FL system at any moment during the training process. As a result, it is critical for the FL system to keep track of malicious clients that leave and then rejoin the system.  However, existing research works do not offer any suggestions for a fix. One of the potential solutions to this issue is to employ either smart contracts or credibility evaluation mechanisms in multiple rounds to identify the reputation of clients in order to keep the adversaries traceable. 

\section{Conclusion}

The emergence of FL is one of the challenging and interesting research directions in the field of machine learning because of in light of rigorous laws for protecting the privacy of participating clients. Nevertheless, the novel concept of FL leads to new challenges due to the invisibility of client-side training data in terms of adversarial attacks, and the poisoning attack is one of such attacks. Recent years have seen an increase in the literature on poisoning attacks and defense strategies to mitigate these attacks. This paper presents a comprehensive summary of current poisoning attacks and defense strategies for FL. Furthermore, we perform a comparative analysis of the state-of-the-art for both poisoning attack and defense in various aspects.  Additionally, the results of experimental evaluations are presented for quantitative comparisons. Finally, we conclude that the study of FL threats is ongoing by pointing out the challenges and future research direction.

\bibliographystyle{IEEEtran}
\bibliography{references}

\end{document}